\documentclass[prb,twocolumn,showpacs,amsmath,amssymb]{revtex4}
\usepackage[dvips]{graphicx}
\usepackage{dcolumn}
\usepackage{bm}

\begin{document}

\title{Weak-field Hall effect and static polarizability of Bloch electrons}

\author{Pavel St\v{r}eda$^{1}$, and Thibaut Jonckheere$^{2}$}

\affiliation{$^{1}$Institute of Physics, Academy of Sciences of the
Czech Republic, Cukrovarnick\'{a} 10, CZ - 162 53 Praha \\
$^{2}$Centre de Physique Th\'{e}orique, Campus de Luminy, case 907, 
F - 13288 Marseille}

\date{\today}

\begin{abstract}
A theory of the weak field Hall effect of Bloch electrons
based on the analysis of the forces acting on electrons
is presented. It is argued that the electric current
is composed of two contributions, that driven by the
electric field along current flow and the non-dissipative
contribution originated in demagnetization currents.
The Hall resistance as a function of the electron concentration
for the tight-binding model of a crystal with square lattice
and body-centered cubic lattice is described in detail.
For comparison the effect of strong magnetic fields is also
discussed.
\end{abstract}


\maketitle

\section{Introduction}

Standard linear response theories for electronic transport
are formulated to obtain the conductivity tensor. Some models
for scattering are needed to get a finite response. The
most popular approximation is the elastic scattering
approach, despite the fact that it cannot compensate
the electron acceleration due to the applied electric field.
It can only be compensated by a momentum dissipation, which
can ensure that the total force acting
on electrons vanishes in the steady state.
Another possibility is to analyze forces acting on electrons
in the transport regime. The condition of the vanishing total force
is the basic physical condition for the steady state. The application
of the linear response approach, limiting the problem to
the case of small deviation from the equilibrium,
gives the transport coefficients satisfying the steady state condition.
This idea will be used to determine the Hall resistance of a system of
Bloch spinless electrons in the case
of a weak magnetic field.
For the sake of simplicity we limit our consideration
to isotropic systems where the energy spectrum is represented
by a single electron band. It can be expected that in
this case the Hall resistance
will not depend on the dissipation explicitly, which can
simplify the analysis substantially.

The Hall resistance is standardly measured on the so-called Hall
bar samples, having the form of a long strip. Far from the
contacts the current density is parallel to the strip edges,
say along $\hat{x}$-direction.
If the magnetic field is applied perpendicularly to the
strip surface, along $\hat{z}$-direction, the current induced
Lorentz force is shifting the electron charge distribution.
As the result there appears a non-equilibrium charge distribution
giving rise the Hall voltage. The steady state requires
the compensation of the Lorentz force by the gradient force.
For isotropic systems this condition can be written in
the following form
\begin{equation}
\frac{B}{c} \, j_x \, = \, - \,
\left\langle \frac{d V(\vec{r})}{d y} \right\rangle \, \equiv \, - \,
\frac{d P(\mu)}{d y}
\; ,
\end{equation}
where $j_x$ is the current density,
$V(\vec{r})$ denotes the background potential including that given by
the electric field along $\hat{y}$ direction, ${\cal{E}}_y$,
and $\mu$ stands for electro-chemical potential.
Angular brackets represent quantum-mechanical and statistical average.
The internal pressure $P(\mu)$ represents the force of the electron
ensemble acting on the external system per unit area.
Linear response approximation with respect to the electro-chemical
potential gradient leads to the following expression for
the Hall resistance
\begin{equation} \label{RH_gen}
R_H \, \equiv \, \frac{U_H}{J} \, = \, - \,
\frac{B}{ec \left(\frac{\partial P(\mu)}{\partial \mu} \right)_B d}
\; ,
\end{equation}
where $J$ is the applied current through the strip of thickness $d$
and $U_H$ denotes the voltage drop between strip edges.

The gradient force and consequently the pressure are
quantities which are generally dependent on the magnetic field
strength $B$. In the weak field limit the Hall resistance can
be supposed a linear function of $B$ which implies that
the $B$-dependent internal pressure can be replaced by
its zero field limit.
For a free electron gas (resp. hole gas) it
can be identified with the so-called statistical pressure,
and its derivative with respect to $\mu$ is simply equal to
the electron concentration $N(\mu)$ (resp. to the negative value of the
hole concentration) \cite{Landau}.  Considering a single electron band
the Hall resistance for chemical potentials in the vicinity
of band edges is thus quite well understood. Since it has opposite
signs at opposite band edges it should vanish at the band
center.
To our knowledge, the only published work in which the transition
between electron to hole like character of the Hall resistance
has been described was based on the application of the Kubo formula
for the special case of substitutional alloys~\cite{Velicky}.
However, no procedure based on force analysis
has been presented. A previous publication~\cite{Streda_wf} 
made by one of us was unfortunately based
on incorrect application of the quasi-classical approach as
will be specified later.

In crystalline solids the equilibrium electron charge 
distribution cannot be assumed as uniform. It is
periodic in real space, having translation symmetry given by the
lattice periodicity. Non-zero current density gives rise to
the Lorentz force inducing a shift of the mass-center positions.
This shift has to be compensated by the gradient force
trying to return it back into the equilibrium distribution.
The results generally depend on the experimental set up,
particularly the way how the non-zero current density is
induced.

In the regime we will call as fully dissipative, the current is
supposed to be exclusively given by the electric field ${\cal{E}}_x$
along strip axis. In other words, if ${\cal{E}}_x \rightarrow 0$
the current vanishes. This assumption requires that
dissipation take place within the strip, i.e. that the system
can be viewed as a macroscopic system. It can be expected
that this fully dissipative regime represents the conditions
of the standardly measured Hall resistance in the weak
field limit for which it can be assumed that the effect of
the magnetic field to the energy spectrum is negligible.

The opposite limiting case is the purely non-dissipative
regime for which the current density is exclusively
determined by the electric field ${\cal{E}}_y$
across the strip, while ${\cal{E}}_x=0$. Such a situation
is observed whenever the Fermi energy is located within the
conduction gaps, i.e. when the magnetic field is strong enough
to induce energy gaps giving rise to a Hofstadter type
spectrum \cite{Hofstadter}.
It has already been shown that in
such a quantum Hall regime the induced Hall current is closely related
to the static electron polarizability \cite{Streda_07,Streda_08}.
The non-dissipative regime, for which the Hall current
is exclusively determined by the orbital magnetization,
can be in principle induced even if the Fermi energy
is located within the energy band. This regime,
in the considered weak field limit, will also be analyzed,
although the resulting effect is expected to be small,
of the order $B^2$.

The main attention will be devoted to the fully dissipative
regime. Vanishing of the total shift of the mass-center position
will be taken as the steady state condition. It will be shown
that it is equivalent to the condition of vanishing
acceleration along the direction perpendicular to both
the current flow and the magnetic field direction.
For the sake of simplicity the outlined idea will
be described in detail for a two-dimensional electron
system since the extension to three-dimensional systems
is straightforward. We will limit our consideration
to the case of a single band given by a square array of tight-binding
atomic states. This model will be described in Section II.
The following Section will be devoted to the determination of
the mass-center shifts within the quasi-classical approach.
The obtained results we will be used to determine
properties of macroscopic systems at zero temperature.
In Section IV the magnetic moment due
to the motion of mass-center positions will be analyzed and
its main features compared with those well known
for the case of a strong magnetic field.
In Section V explicit expressions for
the Hall resistance and the statistical pressure will be derived.
As an example of three-dimensional system the properties
of a body-centered cubic lattice of tight-binding states
will be presented. The Section VI will be devoted to
the non-dissipative regime, which is
closely related to the effect of the magnetic
field on the static electron polarizability. It will be
shown that the polarizability of open systems is modified
by the Lorentz force giving rise to a non-dissipative
Hall current exclusively determined by demagnetization currents.
In the Section VII it will be argued that in standard Hall bar
measurement the current flow is composed of two contributions,
that induced by the electric field along the current flow
and that originated in demagnetization currents. The resulting
general formula for the Hall resistance will be presented
and its properties briefly discussed. The paper will be
closed with short summary.

\section{Zero-field energy spectra}
Tight-binding model is the standard approach to model band 
structure of crystals. If periodic boundary conditions
are applied eigenfunctions are of Bloch form, characterized by
the wave vector $\vec{k}$. Assuming square lattice and
non-zero overlaps between atomic 
functions located at the nearest neighbor atomic 
sites only, the single-band energy spectrum is
\begin{equation}
E^{(0)}(\vec{k}) = - 2 V_0 \,
\left[ \cos(k_1 a) + \cos(k_2 a) \right]
\; ,
\end{equation}
where $a$ and $V_0$ denote lattice constant and overlap integral,
respectively. The components of the wave vector $\vec{k}$
along the main crystallographic axes, (1,0) and (0,1), are $k_1$ and 
$k_2$, respectively. The position of the band center
given by the energy of the atomic orbitals, which can be
represented by a confining frequency $\Omega_0$,
has been chosen as the origin of the energy scale.

\begin{figure}[h]
\includegraphics[angle=0,width=3.3 in]{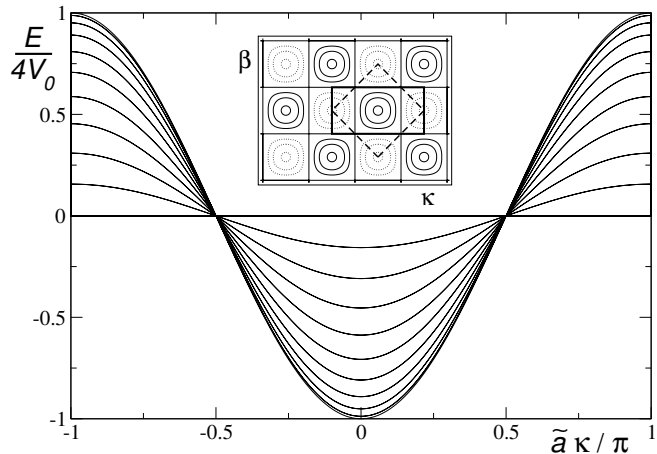}
\caption{Energy dispersions of the tight-binding model for several
fixed values of $\beta \in [-\pi/(2 \tilde{a}),\pi/(2 \tilde{a})]$.
The same dispersions are obtained for fixed values of $\kappa$.
In the inset equi-energy contours are shown in the $\kappa,\beta$ space.
Full and dotted lines correspond to states having electron
or hole like character, respectively. The dashed 
line shows the boundary of the standard first Brillouin zone,
while the bold line shows the elementary cell we are using.}
\label{branches}
\end{figure}

The wave numbers $k_1$ and $k_2$ are not the only choice to
characterize the eigenstates. The square lattice has
a translation symmetry along (1,$\pm$ 1) crystallographic
directions as well. Choosing components of the wave vector
along these directions to characterize Bloch states,
$\kappa \equiv (k_1+k_2)/\sqrt{2}$
and $\beta \equiv (k_1-k_2)/\sqrt{2}$, eigenenergies become
\begin{equation} \label{tb_disp}
E^{(0)}_{\beta}(\kappa) \, = \, - 4 V_0 \,
\cos \left( \kappa \tilde{a} \right) \,
\cos \left( \beta \tilde{a} \right)
\; ,
\end{equation}
where $\tilde{a} = a /\sqrt{2}$. Energy dispersions are
shown in Fig.~\ref{branches}. In the inset the used elementary cell
in the $(\kappa,\beta)$ space, which has been
chosen as $\tilde{a} \beta \in [-\pi/2,\pi/2]$ and
$\tilde{a} \kappa \in [-\pi,\pi]$, is shown.
Since the second derivatives
of the energy with respect of $\beta$ as well as $\kappa$
are equivalent, these states represent quasiparticles having
isotropic effective mass
\begin{equation} \label{m_ast}
\frac{1}{m^{\ast}_{\beta}(\kappa)} = -
\frac{a^2 E_{\beta}^{(0)}(\kappa)}{2 \hbar^2} =
\frac{\cos \left( \beta \tilde{a} \right)}{m^{\ast}_{\rm edge}} \,
\cos \left( \kappa \tilde{a} \right) \,
\; ,
\end{equation}
where $m^{\ast}_{\rm edge}$ denotes the absolute value of the
effective mass at band edges,
$m^{\ast}_{\rm edge}=\hbar^2/(2 V_0 a^2)$.
All quasiparticles of the same energy have the same effective
mass. This property simplifies the quasi-classical approach
which will be described in the following Section.
At negative energies,
$\kappa \tilde{a} \, \epsilon \, (-\pi/2, \pi/2)$,
quasiparticles have electron-like character with positive effective
mass while at positive energies, $\kappa \tilde{a} \, \epsilon \,
\pm (-\pi, -\pi/2)$, quasiparticles have hole-like character
with negative effective mass.

\vspace{7mm}

Electronic transport is studied on samples having the form of a
long strip with a finite width $w$.
It is thus natural to assume periodic boundary conditions along
the direction given by the strip axis only. The eigenfuctions
thus have Bloch-like character along strip axis, while along 
perpendicular direction they are of limited range.
If the strip width can be counted as 
macroscopic, eigenenergies are practically untouched by the
change of the boundary conditions.
For isotropic systems, like the considered square lattice,
the measured transport coefficient are independent of the strip
orientation. We can thus choose the strip axis to be parallel
with the (1,1) crystallographic direction without loss of
generality. In this case the 
eigenfunctions can be approximated by a linear combination
of Bloch states $|\beta, \kappa \rangle$ and
$|-\beta, \kappa \rangle$. The index $\beta$ can be then viewed as
the branch index representing bound modes across the strip.
In such the case the only nonzero component of the velocity
expectation value is along strip axis
\begin{equation} \label{v_0}
v^{(0)}_{\beta}(\kappa) \, = \,
\frac{2 \sqrt{2} V_0 a}{\hbar} \,
\sin(\kappa \tilde{a}) \, \cos(\beta \tilde{a})
\; .
\end{equation}
%

\section{Effective Hamiltonian : quasi-classical approximation}
In the spirit of the preceding Section we consider a strip
opened along the $\hat{x}$ direction which coincides with
(1,1) crystallographic direction of the considered square lattice.
The applied magnetic field along the $\hat{z}$ direction and
the electric field across strip, i.e. along the $\hat{y}$ direction,
${\cal{E}}_y$, give rise to the Lorentz force and electric force,
respectively. To preserve the Bloch character of the wavefunctions
along the $\hat{x}$ direction the Landau gauge for vector
potential is used, $\vec{A} \equiv (-By,0,0)$.
We include it into the Hamiltonian by using the so-called
Peierls substitution \cite{Peierls,Luttinger}.
Since the effective mass of quasiparticles is isotropic
we can use the 
following effective Hamiltonian
\begin{eqnarray} \label{H_eff_0}
H_{\rm eff}^{(\beta,\kappa)}(y) \, = \,
\frac{p_y^2}{2 m_{\beta}^{\ast}(\kappa)} \, + \,
E_{\beta}^{(0)}\left(\kappa - l_B^{-2} y \right) \, +
\nonumber \\ + \,
e {\cal{E}}_y y +
\frac{1}{2} \, m_{\beta}^{\ast}(\kappa) \, \Omega_0^2 y^2 \,
-\frac{\pm 1}{2} \,
\hbar \Omega_0
\; ,
\end{eqnarray}
where the $\pm 1$ corresponds to the sign of the effective mass 
$m^*_{\beta}(\kappa)$,
and $l_B$ denotes the magnetic length, which is related to the
cyclotron frequency $\omega_c$ as
\begin{equation}
l_B^{2} \, \equiv \, \frac{\hbar c}{eB} \, = \,
\frac{\hbar}{m_0 \omega_c}
\; .
\end{equation}
The last term of the effective Hamiltonian $\pm \hbar \Omega_0 /2$
appears to preserve the origin of the energy scale given by the
oscillator energy $\hbar \Omega_0$.
The energy operator $E_{\beta}^{(0)}(\kappa-l_B^{-2}y)$ is
defined by its Taylor expansion in $\kappa$, and
in the considered case of the weak field limit
only terms up to the second order are preserved.
The effective Hamiltonian then becomes simply the one of
an effective harmonic oscillator
\begin{eqnarray} \label{H_eff}
H_{\rm eff}^{(\beta,\kappa)}(y) = 
\frac{p_y^2}{2 m_{\beta}^{\ast}(\kappa)} +
\frac{m_{\beta}^{\ast}(\kappa)}{2} \, \tilde{\Omega}^2
\left[ y - Y_{\beta}(\kappa) \right]^2 +
\nonumber \\ -
\frac{1}{2} \, m_{\beta}^{\ast}(\kappa) \, \tilde{\Omega}^2
\left[ Y_{\beta}(\kappa) \right]^2 \, + \,
E_{\beta}^{(0)}(\kappa) \, - \,
\frac{\pm 1}{2} \, \hbar \Omega_0
\, , \;
\end{eqnarray}
where $Y_{\beta}(\kappa)$ denotes the expectation value
of the mass-center position
\begin{equation} \label{Y_eff}
Y_{\beta}(\kappa) =
\frac{m_0 \omega_c v_{\beta}^{(0)}(\kappa) - 
e {\cal{E}}_y}
{m_{\beta}^{\ast}(\kappa) \tilde{\Omega}^2}
\; ,
\end{equation}
and
\begin{equation}
\tilde{\Omega}^2 \, \equiv \,
\Omega_0^2 + 
\left(\frac{m_0}{m^{\ast}_{\beta}(\kappa)} \right)^2 \omega_c^2
\; .
\end{equation}
Resulting eigenenergies are
\begin{equation}
E_{\beta}(\kappa) =
\pm \hbar \frac{\tilde{\Omega}-\Omega_0 }{2} + E^{(0)}_{\beta}(\kappa) -
\frac{m_{\beta}^{\ast}(\kappa)}{2} \, 
\tilde{\Omega}^2 \,
\left[ Y_{\beta}(\kappa) \right]^2 .
\end{equation}
The expectation of the velocity along the strip axis
has the following expression
\begin{equation} \label{v_B}
v_{\beta}(\kappa) \, = \, v^{(0)}_{\beta}(\kappa) \, - \,
\frac{m_0}{m_{\beta}^{\ast}(\kappa)} \,
\omega_c \, Y_{\beta}(\kappa)
\; ,
\end{equation}
which coincides with the well known result for
magnetic field corrections to the velocity \cite{Kubo}.

The corrections to the energy, as well as to the velocity,
due to the presence of magnetic
and electric fields are proportional to square of these fields
or their product.
In the weak field limit, $B \rightarrow 0$ and
${\cal{E}}_y \rightarrow 0$, these corrections can thus be neglected
and the only effect that will be considered is the change of the
quasiparticle dynamics, represented in our description by
the change of their mass-center positions. This approach which will be
used in the following treatment is in accord with
the standard quasi-classical view. To support this let us
consider the product $\Omega_0^2 Y_{\beta}(\kappa)$
representing the quasiparticle acceleration $a_y(\beta,\kappa)$
along $\hat{y}$ direction. From Eq.~(\ref{Y_eff}) we get
\begin{equation} \label{acceleration}
a_y(\beta,\kappa) \, = \,
\frac{m_0 \omega_c v_{\beta}^{(0)}(\kappa) -
e {\cal{E}}_y} 
{m_{\beta}^{\ast}(\kappa)}
\; .
\end{equation}
This relation leads to the conclusion that in crystals the
acceleration of quasiparticles is modified
by their effective mass which is in agreement
with the quasi-classical approach presented e.g. in
Landau-Lifshitz textbook \cite{Landau}.
Note that the quasiparticle acceleration along the $\hat{y}$ direction
induced by the electric field ${\cal{E}}_y$ originates in their
transfer between branches $\beta$.

The presented quasi-classical approach neglects the interference
effects induced by the magnetic field which are responsible
for modification of the energy spectrum. A number of
energy gaps are created and the energy structure is
of the Hofstadter type \cite{Hofstadter}.
In weak field limit, the gaps in the energy spectrum becomes
extremely small, and it can be expected that theses interferences
will be destroyed by dissipative processes 
always present at finite temperatures. For this reason
the presented quasi-classical approach is acceptable.

\section{Magnetic moment of Fermi electrons}
The applied magnetic field along the $\hat{z}$ direction gives
rise to the magnetic moment $M_z(\mu)$. Generally $M_z(\mu)$ can be
divided into two contributions: $M_z^{(i)}(\mu)$ given
by the internal momentum of quasiparticles and $M_z^{(a)}(\mu)$
due to the motion of their mass-centers. The second
contribution can be viewed as the macroscopic one since
the trajectories of the mass-center positions are extended
along the $\hat{x}$ direction. It can easily be determined
within the quasi-classical approximation presented in
the preceding Section. Its expectation value reads
\begin{equation}
M_z^{(a)}(\beta,\kappa) \, = \, \frac{e}{c} \,
v_{\beta}^{(0)}(\kappa) \, Y_{\beta}^{(B)}(\kappa)
\; ,
\end{equation}
where in accord with Eq.~(\ref{Y_eff})
\begin{equation} \label{Y_B}
Y_{\beta}^{(B)}(\kappa) \, = \,
\frac{m_0 \omega_c v_{\beta}^{(0)}(\kappa)}
{m_{\beta}^{\ast}(\kappa) \Omega_0^2}
\; .
\end{equation}
Defining the dimensionless quantity $\tilde{\sigma}(\mu)$ as
\begin{equation} \label{sigma_gen}
\tilde{\sigma}(\mu) \equiv \frac{2 \pi}{w}
\sum_{\beta,\kappa}
f_0 \left(E_{\beta}^{(0)}(\kappa) - \mu \right)
\frac{d Y_{\beta}^{(B)}(\kappa)}{d \kappa}
\; ,
\end{equation}
where $w$ is the strip width,
the Fermi electron contribution per unit area reads
\begin{equation} \label{M_z}
\left( \frac{\partial M_z^{(a)}(\mu)}{\partial \mu} \right)_{B}
\, = \,
\frac{e}{h c} \, \tilde{\sigma}(\mu)
\; .
\end{equation}

The relation between $M_z^{(a)}(\mu)$ and $\tilde{\sigma}(\mu)$
defined by Eq.~(\ref{sigma_gen}) is quite general and for
the case of quantizing magnetic fields and weak periodic
modulation has already been discussed in detail, where
$\tilde{\sigma}(\mu)$ was called
the effective topological number \cite{Streda_08,Streda_07}.

\begin{figure}[h]
\includegraphics[angle=0,width=3.3 in]{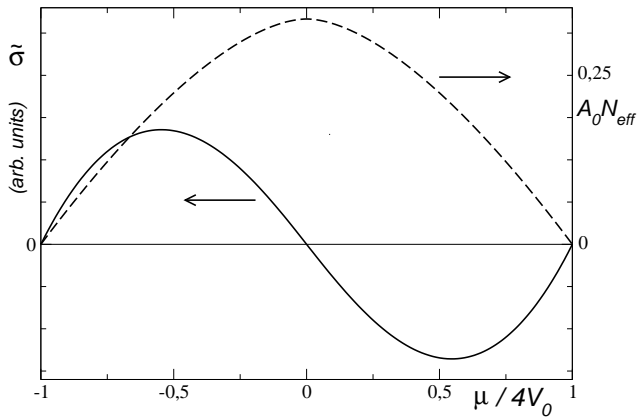}
\caption{Energy dependence of the effective topological
number $\tilde{\sigma}(\mu)$, full line, and the effective
quasiparticle number in a unit cell $A_0 N_{\rm eff}(\mu)$,
dashed line.}
\label{sigma_E}
\end{figure}

In the weak field limit, for the square lattice,
performing the $\kappa$ integration in Eq.~(\ref{sigma_gen})
gives
\begin{equation}
\tilde{\sigma}(\mu) = \frac{1}{w} \sum_{\beta}
2 Y^{(B)}_{\beta}(\kappa^{\beta}_F)
\end{equation}
where $\kappa^{\beta}_F$ if the positive value of $\kappa$
on the Fermi surface for a given $\beta$. Using 
$1/m^*_{\beta}(\kappa^{\beta}_F) = -\mu/(4 V_0 m^*_{edge})$,
we finally get
\begin{equation} \label{sigma_exp}
\tilde{\sigma}(\mu) \, = \, - \,
\frac{m_0}{m^{\ast}_{\rm edge}} \, \frac{\omega_c}{\Omega_0} \,
\frac{\mu}{\hbar \Omega_0} \, \pi a^2 \, N_{\rm eff}(\mu)
\; ,
\end{equation}
where
\begin{eqnarray} \label{N_eff}
N_{\rm eff}(\mu) \, \equiv \,
\frac{2}{\pi^2 a^2}
\int_0^{\phi_0} \sqrt{\cos^2 \phi - \cos^2 \phi_0} \, d \phi
\; , \nonumber \\
\phi_0={\rm arccos} (|\mu|/4V_0)
\; .
\end{eqnarray}
At the band edges the effective quasiparticle
concentration $N_{\rm eff}(\mu)$ approaches the electron or
hole concentration $m^*_{edge} /(2 \pi \hbar^2) \Delta E$
with $\Delta E = \mu + 4 V_0$ for electrons and
$\Delta E = 4 V_0 - \mu$ for holes. 
Energy dependence of $\tilde{\sigma}(\mu)$ and $N_{\rm eff}(\mu)$
are shown in Fig.~\ref{sigma_E}.

\vspace{7.0mm}

At negative energies, $\mu < 0$, for which quasiparticles have 
electron-like
character, the Fermi contribution to $M_z^{(a)}(\mu)$ given
by Eq.~(\ref{M_z}) is positive, i.e. it represents
paramagnetic contribution to the total magnetic moment.
It is often interpreted as a contribution given by electrons
skipping along sample edges, which fully compensates the
diamagnetic moment of electrons in the classical limit.
In the presented model it has been equally splited over
the local strips of tight-binding atomic orbitals.
At positive energies the contribution of Fermi particles
to $M_z^{(a)}(\mu)$ has opposite sign revealing
electron-hole symmetry. At the band center it vanishes,
$\tilde{\sigma}(\mu) \rightarrow 0$. This is the consequence
of the electron-hole symmetry and this feature persists
even in the case of strong magnetic fields affecting
energy spectra substantially \cite{Streda_08}.
Similarly, when the energy band is fully occupied,
$\tilde{\sigma}(\mu) \rightarrow 0$ in the weak field limit,
as well as in the case of quantizing magnetic fields.

\section{Hall resistance: fully dissipative regime}
\subsection{Square lattice}
The electric field along the strip axis, ${\cal{E}}_x$, accelerates
electrons along the $\hat{x}$ direction. This effect can be
modelled by including the time dependent vector potential
$\vec{A}(t) \equiv (-e {\cal{E}}_x t,0,0)$ into the
Hamiltonian. For small values of the electric field and short times
the linear response gives the following change of the
quasiparticle velocity
\begin{equation}
v_{\beta}^{(0)}(\kappa-e {\cal{E}}_x t/\hbar) \, \approx \,
v_{\beta}^{(0)}(\kappa) \, - \,
\frac{d v_{\beta}^{(0)}(\kappa)}{ d \kappa} \,
\frac{e{\cal{E}}_x t}{\hbar}
\; ,
\end{equation}
where corrections to the velocity proportional to $B^2$,
Eq.~(\ref{v_B}), have been neglected.
Summation over occupied states gives the current density along
$\hat{x}$ direction
\begin{equation} \label{j_x}
j_x(\mu) \, = \,
\frac{eN_{\rm eff}(\mu)}{m^{\ast}_{\rm edge}} \,
e {\cal{E}}_x t
\; .
\end{equation}

Change of the velocities gives rise to the shift of mass-center
positions of quasiparticles, Eq.~(\ref{Y_B}). Summation over
occupied states leads to the following expression for
mass-center shift $\langle \Delta Y^{(B)}(\mu) \rangle$
of the electron density within the unit cell area 
\begin{equation}
\frac{\langle \Delta Y^{(B)}(\mu) \rangle}{A_0} \,  = \, -
\frac{\tilde{\sigma}(\mu)}{2 \pi} \, 
\frac{e{\cal{E}}_x t}{\hbar} \, = \,
\frac{B j_x}{c \, m^{\ast}_{\rm edge} \Omega_0^2} \,
\frac{\mu}{4 V_0}
\, .
\end{equation}
Evidently this time dependent shift is induced by the
Lorentz force.

\begin{figure}[h]
\includegraphics[angle=0,width=3.3 in]{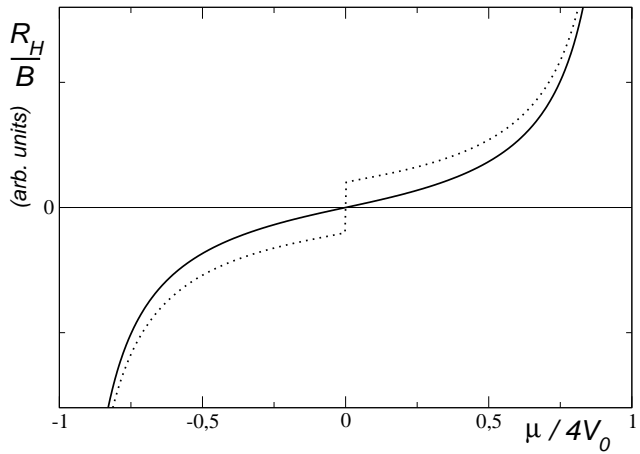}
\caption{Energy dependence of the Hall resistance for a two-dimensional
square lattice. Dotted line represents the classical resistance for free
carriers, i.e. the quantity $4 V_0 N_{\rm eff}(\mu)/\mu$
is replaced by the negative value of the electron
concentration ($\mu <0$) or by the hole concentration ($\mu >0$).}
\label{RH_E}
\end{figure}

The shift of the mass-center position is closely related to
the shift of the electron charge distribution with respect
of the periodic positive background charge. It gives rise
to the Coulomb energy. For the system is thus energetically
more acceptable to induce electric field along $\hat{y}$
direction, ${\cal{E}}_y$, which would be able to minimize
the Coulomb energy, i.e. to force shifted electron charge
distribution towards its equilibrium one. In the presented
model this force is represented by the confining potential
of the strength given by the frequency $\Omega_0$. 
Standard condition to estimate the induced field ${\cal{E}}_y$
is the condition of vanishing acceleration given by
Eq.~(\ref{acceleration}). It coincides with the condition
of vanishing average shift of mass center positions defined
by Eq.~(\ref{Y_eff}).
Summation over occupied states gives
\begin{equation}
\frac{B j_x}{c m^{\ast}_{\rm edge}}
\frac{\mu}{4 V_0} \, - \,
e {\cal{E}}_y \frac{N_{\rm eff}(\mu)}{m^{\ast}_{\rm edge}}
\, = \, 0
\, .
\end{equation}
For the Hall resistance we get
\begin{equation} \label{RH_2D}
R_H(\mu) \, \equiv \, \frac{{\cal{E}}_y}{j_x} \, = \,
\frac{B}{ec N_{\rm eff}(\mu) \, \frac{4 V_0}{\mu}}
\, .
\end{equation}
The energy dependence of the Hall resistance is shown
on Fig.~\ref{RH_E}. It shows clearly the electron hole symmetry.
At band edges, $\mu \rightarrow \pm 4 V_0$, $N_{\rm eff}(\mu)$
approaches quasiparticle concentration and the classical result
for the Hall resistance is recovered.

\vspace{7.0mm}

To approach a steady state an
energy dissipation is necessary to eliminate the acceleration
induced by ${\cal{E}}_x$. The standard way is to assume
exponential decay of momentum characterized by the relaxation
time $\tau$. In other words the time entering expression for
the current density, Eq.~(\ref{j_x}), has to be replaced by
$\tau$, which is in general a $\mu$ dependent quantity.
Although the Hall resistance does not explicitly depends
on the momentum
relaxation it is essential for application of the linear
response with respect of ${\cal{E}}_y$ to held it
sufficiently small.

In the considered weak field limit the current is supposed
to be fully
determined by the electric field along the strip axis. This
assumption is applicable if the dissipation take place
within the strip area. In the case of a quantizing
magnetic field this assumption is not acceptable.
Whenever the Fermi energy is located within
an energy gap there might appear magnetic edge states leading
to non-dissipative current. In this case the electric field along
the current flow vanishes and the current is fully
determined by the perpendicular electric field.

\subsection{Statistical pressure and the Hall resistance}
Comparison of the expression for the Hall resistance,
Eq.~(\ref{RH_2D}), with the more general form given by the
Eq.~(\ref{RH_gen}) suggests that Hall resistance has to be
related to a pressure gradient representing the fully dissipative
regime.

In the quasi-classical approach we are using, the carriers are supposed
to be quasiparticles having
isotropic effective mass $m_{\beta}^{\ast}(\kappa)$ which is
defined by their energy, Eq.~(\ref{m_ast}). Quasiparticles
are allowed to move freely along any direction (i.e. even along $\hat{y}$
direction by transitions between branches). The effect of the periodic 
background is included via their effective mass and
their momentum is given by the product of the effective
mass and the velocity.
In this case the so-called statistical pressure can
be easily evaluated following the
standard procedure for gas system \cite{Landau}.
Let us consider quasiparticles located within a finite area
separated from outside quasiparticles by walls preventing
particle transfer. The total
momentum transfered to the wall perpendicular e.g. to
$\hat{x}$ direction is equal to
$2m_{\beta}^{\ast}(\kappa) v_{\beta}^{(0)}(\kappa)$.
The number of electrons hitting the wall per unit time
is given by their velocity, and statistical
pressure defined as the force per unit length acting along
any direction reads
\begin{equation} \label{P_stat}
P^{\rm (st)}(\mu) = 2 
{\sum_{\beta, \kappa}}'
f_0 \left( E_{\beta}^{(0)}(\kappa)-\mu \right) \,
m_{\beta}^{\ast}(\kappa) \left[ v_{\beta}^{(0)}(\kappa) \right]^2
\, ,
\end{equation}
where the prime indicates that summation is taken over those
states for which the velocity is of one particular sign,
say the positive one.
Inserting explicit expressions for the effective
mass and velocity, Eq.~(\ref{m_ast}) and Eq.~(\ref{v_0}),
respectively, we get
\begin{equation} \label{dP_stat/dmu}
\left( \frac{\partial P^{\rm (st)}(\mu)}{\partial \mu} \right)_{B=0}
\, = \, - \,
\frac{4 V_0}{\mu} \, N_{\rm eff}(\mu)
\; .
\end{equation}
Contribution of Fermi carriers to the statistical pressure is
positive for electron-like quasiparticles while for hole-like
quasiparticles it is negative since the hole concentration
decreases with rising value of the chemical potential $\mu$.
At the band center where the density of states diverges, the
statistical pressure diverges as well and consequently the
Hall resistance vanishes.

Comparison of Eq.~(\ref{dP_stat/dmu}) with Eq.~(\ref{RH_2D})
gives for the Hall resistance in the fully dissipative regime
the relation we have been looking for
\begin{equation} \label{RH_Pstat}
R_H(\mu) \, = \, - \,
\frac{B}{ec
\left( \frac{\partial P^{\rm (stat)}(\mu)}{\partial \mu} 
\right)_{B=0}}
\, .
\end{equation}
Note that for two-dimensional systems the strip thickness $d$
entering the general formula, Eq.~(\ref{RH_gen}), has to be replaced
by unity.

The above calculation corrects a previous result presented by
one of us~\cite{Streda_wf}, where the
effective mass was erroneously identified with the local
cyclotron mass.

\subsection{Body-centered cubic lattice}
The results presented above can easily be generalized to a
three dimensional system. As an example we consider here the
body-centered cubic lattice.
Taking into account the overlap
between nearest neighbor atomic sites only,
the tight-binding single band spectrum, in the analogy with
that for square lattice, can be written in the following form
\begin{equation}
E_{\beta,\gamma}^{(0)}(\kappa) \, = \, - \, 8 V_0 \,
\cos(\kappa a) \, \cos(\beta a) \, \cos(\gamma a)
\; ,
\end{equation}
where index $\gamma$ represents eigenstate modes along
$\hat{z}$ direction. The effective mass is isotropic and
all states of the same energy have the same effective mass 
\begin{equation}
\frac{1}{m_{\beta,\gamma}^{\ast}(\kappa)} \, = \, - \,
\frac{a^2 E_{\beta,\gamma}^{(0)}(\kappa)}{\hbar^2}
\; .
\end{equation}
For negative energies particles have electron-like character
while for positive energies they have hole-like character.
At the band edges the absolute value of the effective mass
is $m_{\rm eff}^{\ast}= \hbar^2/(8 V_0 a^2)$. We can thus proceed
as in the preceding section. We get expressions
with the same structure, but with an additional summation over the index
$\gamma$. For the Hall resistance
defined by Eq.~(\ref{RH_gen}) we get
\begin{equation} \label{RH_bcc}
R_H^{\rm (bcc)}(\mu) \, = \,
\frac{B}{ec N_{\rm eff}^{\rm (bcc)}(\mu) \, \frac{8 V_0}{\mu} \, d}
\, ,
\end{equation}
where
\begin{eqnarray}
N_{\rm eff}^{\rm (bcc)}(\mu) \, =  \, \frac{4}{\pi^3 a^3}
\int_0^{\phi_0} d \phi \times
\qquad \qquad \qquad \qquad \nonumber \\ \times
\int_0^{\theta_{\phi}}
\sqrt{\cos^2 \phi \, \cos^2 \theta - \cos^2 \phi_0} \, d \theta
\, ,
\end{eqnarray}
$\theta_{\phi}={\rm arccos}[|\mu|/(8V_0 \cos \phi)]$.
In this case also the condition of vanishing acceleration
as well as the independently derived expression for the statistical
pressure lead to the same results.
At the lower band edge ($\mu \rightarrow -8V_0$)
$N_{\rm eff}^{\rm (bcc)}(\mu)$ approaches
the electron concentration, while at the upper edge
($\mu \rightarrow 8V_0$) it approaches the hole concentration.
At the band center the Hall resistance vanishes as expected.
As function of the energy it shows the same
qualitative features as that for square lattice presented
in Fig.~\ref{RH_E}.

\section{Effect of the magnetic field on electron polarizability}
It has recently been shown that a strong magnetic field,
which leads to magnetic quantization of the electron energy spectra,
can significantly affect the static electron polarizability of crystalline 
solids. This is caused by appearance of the Lorentz force.
As a result a close relation between the induced
Hall current and the static electron polarizability of
two-dimensional
systems open along one direction has been established
\cite{Streda_07,Streda_08}. In the weak field limit we consider here,
it can be
expected that this effect will be nearly negligible since
the effect of the magnetic field on the energy spectra is
proportional to square of the field strength. Nevertheless,
the analysis of this effect will allow us to
understand the difference between this purely non-dissipative
regime and the fully dissipative one.

Let us consider the same geometry as that used for 
discussion of the Hall resistance in fully dissipative
regime, i.e. a strip of the
square lattice of tight-binding atomic states opened along
$\hat{x}$ direction with electron concentration $N(\mu)$.
To establish the electron polarizability the strip has to
be placed between capacitor plates.
To model non-dissipative process the charging of the plates has
to be slow to allow adiabatic evolution of the electron system
within the strip: at any time the electrons are  in a
quantum eigenstate. No current across strip is allowed,
i.e. contrary to fully dissipate regime electron transitions
between energy branches are forbidden.
The resulting charge density redistribution
across the strip, accompanied by an internal electric
field ${\cal{E}}_y$, more precisely by the gradient of the
electro-chemical potential, can be characterized 
by the shift of the mass-center positions.
Although ${\cal{E}}_y$ is not uniform
across the strip, within the linear response 
approach the mass-center shift can be split into the
local shifts per unit cell induced by an average
field ${\cal{E}}_y$.

To estimate the static electron polarizability we follow the same
procedure as that already used for the case of quantizing
magnetic field~\cite{Streda_07,Streda_08}.
Let us first discuss the case of zero magnetic field.
For small deviations from the equilibrium allowing linear
response approach, the condition
of the vanishing total force, defining the shift of the
mass-center position within the unit cell area, reads
\begin{equation} \label{dY_0}
- m_0 \Omega_0^2 \Delta Y - e \,
A_0 N(\mu) \, {\cal{E}}_y \, = \, 0
\; ,
\end{equation}
where $A_0=a^2$ denotes unit cell area. The force
$- m_0 \Omega_0^2 \Delta Y$ represents harmonic approximation
of the gradient force originated in the shift of the electron
charge with respect of the background positive charge distribution.
The static electron polarizability $\alpha^{(0)}(\mu)$ is defined
as the total dipole moment per unit area divided by the electric
field \cite{Ashcroft} and we get
\begin{equation} \label{alpha_0}
\alpha^{(0)}(\mu) \, = \,
\frac{e^2}{m_0 \Omega_0^2} \, N(\mu)
\; .
\end{equation}
Note that the confining frequency $\Omega_0$ corresponds to that
determining the energy of atomic states only in the
limit of vanishing overlap, $V_0 \rightarrow 0$.
Generally it depends on the electron concentration as well as
on the overlap strength. It has thus be viewed as the parameter
depending on the chemical potential. The same is true for
the confining frequency entering effective Hamiltonian,
Eq.~(\ref{H_eff_0}).

Electric field gives rise to a shift of atomic orbitals defined by
Eq.~(\ref{dY_0}), $-e {\cal{E}}_y/(m_0 \Omega_0^2)$
along $\hat{y}$ direction. In the presence of the magnetic
field there appears additional effect given by
change of the vector potential value at the center of
shifted atomic orbitals. It can be estimated by the Peierls 
substitution leading to the shift of the wave number $\kappa$
\begin{equation}
\kappa \, \rightarrow \, \kappa \, + \, \Delta \kappa
\quad , \quad
\Delta \kappa \, \equiv \, \frac{\omega_c^2}{\Omega_0^2} \,
\frac{e {\cal{E}}_y}{\hbar \omega_c}
\; .
\end{equation}
In the weak field limit the expansion up to the second order
in the magnetic field strength
gives the following shift of the mass-center position
\begin{equation} \label{Delta_Y_B}
\Delta Y_{\beta}(\kappa) \, = \, - \,
\frac{e {\cal{E}}_y}{m_0 \Omega_0^2}
\left(
1- l_B^{-2} \, \frac{d Y_{\beta}^{(B)}(\kappa)}{d \kappa}
\right)
\; .
\end{equation}
For the average 
mass-center shift $\langle \Delta Y(\mu) \rangle$
of the electron density within the unit cell area induced by
the electric field ${\cal{E}}_y$  we get
\begin{equation} \label{Delta_Y}
\langle \Delta Y(\mu) \rangle \, = \, - \, A_0 \,
\frac{e {\cal{E}}_y}{m_0 \Omega_0^2}
\left( N(\mu) - \frac{\tilde{\sigma}(\mu)}{2 \pi l_B^{2}}
\right)
\; .
\end{equation}

The mass-center shifts give rise a current along the $\hat{x}$
direction, the induced Hall current.
Within linear response with respect to the electric field
the energy becomes dependent on the mass-center position
\begin{equation}
E_{n,\beta}(\kappa, {\cal{E}}_y) \, = \,
E_{\beta}^{(0)}(\kappa) + e {\cal{E}}_y \,
\left( n \tilde{a} + Y_{\beta}^{(B)}(\kappa) \right)
\; ,
\end{equation}
where $n \tilde{a}$ is the position of $n$-th local strip.
The resulting change of the velocity
\begin{equation}
\Delta v_{\beta}(\kappa) \, = \,
\frac{e {\cal{E}}_y}{\hbar} \,
\frac{d Y_{\beta}^{(B)}(\kappa)}{d \kappa}
\; ,
\end{equation}
leads to the following expression for the induced Hall current
density
\begin{equation} \label{j_H}
j_H(\mu) \, = \, - \,
\frac{e^2}{h} \, \tilde{\sigma}(\mu) \, {\cal{E}}_y
\; .
\end{equation}

The equality given by Eq.~(\ref{Delta_Y}) is the consequence of the
condition of the vanishing total force acting on electrons
which reads
\begin{equation}
- \, \frac{1}{A_0} \, m_0 \Omega_0^2 \langle \Delta Y(\mu) \rangle - 
e {\cal{E}}_y N(\mu) - 
\frac{B}{c} \, j_H \, = \, 0
\; .
\end{equation}
In comparison with the zero field case (Eq.~(\ref{dY_0})),
the presence of the magnetic field gives rise to another
term, the Lorentz force.

For the static electron polarizability
$\alpha(\mu)$ we get
\begin{equation}
\alpha(\mu) =
\frac{e^2}{m_0 \Omega_0^2}
\left( N(\mu) -
\frac{\tilde{\sigma}(\mu)}{2 \pi l_B^2} \right) \equiv
\frac{e^2}{m_0 \Omega_0^2} \, \tilde{s}(\mu)
\, ,
\end{equation}
where the last equality defines $\tilde{s}(\mu)$.

As in the case of a quantizing magnetic field~\cite{Streda_08}, 
the corrections are due to the existence of macroscopic demagnetization
currents responsible for non-zero value of $\tilde{\sigma}(\mu)$
as follows from Eq.~(\ref{M_z}).
In contrast to the case of a quantizing magnetic field, the
magnetic corrections to the static electron polarizability are
practically negligible in the weak field limit since they are
proportional to $B^2$.

Finally, note that we have analyzed the effect of the magnetic
field to the
polarizability of a strip opened along $\hat{x}$ direction,
which models a strip of finite length with periodic
boundary conditions. The results are thus applicable
also for Corbino samples of large radius, placed between
cylindrical capacitor plates, i.e. a system which can
be experimentally realized.

\section{General Hall resistance formula}
Two origins of the current induced in the open strip have been
discussed. First, the current induced by an electric
field ${\cal{E}}_x$ applied along strip axis, given by
Eq.~(\ref{j_x}). In this case, we call as fully dissipative,
electric field ${\cal{E}}_y$ across the strip has been introduced
to fully compensate acceleration of electrons along
$\hat{y}$ direction induced by ${\cal{E}}_x$. By another words
the field ${\cal{E}}_y$ was supposed to return charge distribution
across the strip back into its equilibrium one.
Second, assuming zero electric field along strip axis,
the non-dissipative current $j_H$ induced by an
electric field ${\cal{E}}_y$ across the strip,
Eq.~(\ref{j_H}), has been established. This field, ${\cal{E}}_y$,
gives rise to the non-equilibrium charge distribution modeled by
the shift of the equilibrium distribution. Resulting current
density $j_H$ originates in the response of macroscopic 
demagnetization currents to the electric field ${\cal{E}}_y$.

However, the condition defining fully dissipative regime is
not realistic in principle. Electric field ${\cal{E}}_y$
cannot exists without shift of the charge distribution
across the strip which gives rise the non-zero current density
$j_H$. Within linear response with respect of electric fields,
${\cal{E}}_x$ and ${\cal{E}}_y$, the current density
is thus given by the sum of both contributions,
$j = j_x+j_H$. Consequently, for the Hall resistance
we get   
\begin{equation} \label{R_H_tot}
R_H(\mu) \equiv \frac{{\cal{E}}_y}{j} = -
\frac{B}{ec \left[
\left(
\frac{\partial P^{\rm (st)}(\mu)}{\partial \mu} \right)_B
+ \frac{\tilde{\sigma}(\mu)}{2 \pi l_B^2} \right] d}
\, .
\end{equation}
Comparison with its general form, Eq.~(\ref{RH_gen}),
and the use of the relation between $\tilde{\sigma}(\mu)$ and
macroscopic part of the magnetic moment, Eq.~(\ref{M_z}),
give the following expression for the contribution
of Fermi electrons to the internal pressure
\begin{equation}
\left(\frac{\partial P(\mu)}{\partial \mu} \right)_B =
\left(
\frac{\partial P^{\rm (st)}(\mu)}{\partial \mu}
\right)_B + B
\left( \frac{\partial M_z^{(a)}(\mu)}{\partial \mu} \right)_{B}
 .
\end{equation}
In the considered quasi-classical approach the internal
pressure is thus composed of two contributions, the
statistical pressure and that induced by the magnetic field
$\vec{B} \cdot \vec{M}^{(a)}$.
In the weak field limit the correction term
$\tilde{\sigma}(\mu)/(2 \pi l_B^2)$ is proportional to
$\omega_c^2/\Omega_0^2 \sim B^2$ and can thus be neglected.

The expression for the Hall resistance given by Eq.~(\ref{R_H_tot})
is applicable to the case of strong quantizing magnetic fields
as well. For two-dimensional systems the single band energy
spectrum is split into magnetic subbands for which a quasi-classical
approach describing quasiparticle dynamics can be developed.
However, the resulting statistical pressure will be a magnetic 
field dependent quantity. 
The corresponding non-dissipative currents
have already been analyzed in detail and the properties
of the effective topological number
$\tilde{\sigma}(\mu)$ well understood \cite{Streda_07,Streda_08}.
For fully occupied magnetic subbands the derivative of the statistical
pressure with respect of the chemical potential vanishes,
$\tilde{\sigma}(\mu)$ approaches an integer value, and
the quantum Hall resistance is recovered~\cite{Streda_wf}.

\section{Summary}
We have applied a quasi-classical approach to establish the Hall
resistance of Bloch electrons in the weak field limit.
The single tight-binding band for a square lattice and for a body
centered cubic lattice have been used as model systems. In both
cases quasiparticles having an isotropic effective mass
can be introduced which simplifies the description significantly.
To obtain the Hall resistance the forces acting on the quasiparticles
have been analyzed. The resulting dependence of the Hall resistance
on the Fermi energy, i.e. on the electron concentration,
shows a smooth transition from electron to hole like
character. It has zero value at the band center as expected. 

The role of macroscopic demagnetization currents,
often treated as non-dissipative edge currents,
has also been analyzed and their effect
to the Hall resistance established. While
in the weak field limit their contribution can be
neglected, in quantizing magnetic fields they are responsible
for the quantum Hall effect.

\section*{Acknowledgment}

This research was supported by the Grant Agency of the
Czech Republic under Grant No. 202/08/0551 and by the
Institutional Research Plan No. AV0Z10100521.
P. St\v{r}eda acknowledges support of CPT (UMR6207 of CNRS)
and the Universite Sud Toulon Var for their hospitality.

\end{document}